\documentstyle[prl,aps,twocolumn,floats,draft]{revtex}

\newcommand{\mywidetext}[1]{\twocolumn[
\hsize\textwidth\columnwidth\hsize\csname@twocolumnfalse\endcsname
#1 ] }

\begin{document}
\input epsf.sty
\input psfig.sty

\newcommand{\nn}{\nonumber \\}
\newcommand{\be}{\begin{equation}}
\newcommand{\ee}{\end{equation}}

\newcommand{\chem}[1]{$\mathrm{#1}$}
\newcommand{\MgB}{\chem{MgB_2}}
\newcommand{\K}{{\mathrm{K}}}
\newcommand{\T}{{\mathrm{T}\,}}
\newcommand{\Ang}{{\mathrm{\AA}}}
\newcommand{\eV}{{\mathrm{eV}}}

\newcommand{\rvec}{{\mathbf{r}}}
\newcommand{\qvec}{{\mathbf{q}}}
\newcommand{\kvec}{{\mathbf{k}}}
\newcommand{\rpa}{{\mathrm{rpa}}}

\newcommand{\va}{v_a}
\newcommand{\latticea}{3.086\Ang}
\newcommand{\latticec}{3.524\Ang}

\mywidetext{
  
  \title{Acoustic Plasmons in \MgB} 
  
  \author{K.~Voelker$\,^1$, V.~I.~Anisimov$\,^{1,2}$, and T.~M.~Rice$\,^1$}
  \address{$^1$Theoretische Physik, ETH H\"onggerberg, CH-8093 Z\"urich,
    Switzerland} \address{$^2$Institute of Metal Physics, Russian Academy of
    Sciences, Ekaterinburg, Russia} \date{\today}
  
  \maketitle
  
  \begin{abstract}
    We present strong evidence for the existence of an acoustic plasmon mode,
    that is, a quadrupolar charge collective mode with linear dispersion,
    in \MgB. This mode may be responsible for the anomalously small value of
    the Coulomb pseudopotential required to explain the high
    superconducting transition temperature of $40\K$.
  \end{abstract}
  
  \pacs{}
  
  }


The recent discovery of superconductivity in~\MgB~\cite{Akimitsu} has spurred
a wave of activity both on the experimental and on the theoretical side. The
transition temperature of $T_c = 39.5\K$ is by far the highest found in any
binary compound. With the exception of $\mathrm{C_{60}}$, it is surpassed only
by the high-temperature copper-oxide superconductors.  A detailed analysis of
the electron-phonon interaction finds relatively strong coupling to a high
frequency phonon mode \cite{Andersen}. However, the observed $T_c = 39.5\K$
could be obtained only by assuming that the Coulomb pseudopotential takes an
anomalously small value, $\mu^*\simeq 0.02$ --- a value roughly seven times
smaller than for usual $s$-$p$ superconductors. Here, we propose that an
acoustic plasmon mode exists in \MgB. This leads to an attractive
contribution, thereby lowering the value of the Coulomb
pseudopotential.
  
The idea of acoustic plasmons goes back to Pines \cite{Pines}, and, following
Fr\"ohlich, has been considered as a mechanism for superconductivity in the
transition metals in the past \cite{Froehlich}.  In the presence of two
carrier species with very different effective masses, the light carriers can
act to screen the Coulomb repulsion between the heavy carriers, and thereby
lower the plasma frequency of the heavy carriers. The resulting plasmon mode
will have a linear dispersion $\omega_a = \va q$ (hence the term acoustic
plasmon), where $\va$ is of the order of (but somewhat higher than) the Fermi
velocity of the heavy carriers.  The acoustic plasmon mode is Landau damped by
decay into the electron-hole continuum of the light carriers, but for small
wave vectors $q$ the mode will be located within the lower range of the
electron-hole continuum, so that the damping may be weak, and the acoustic
plasmon may exist as a well-defined collective mode.
 
Since no experimental evidence of acoustic plasmons could be found, and one
was able to explain superconductivity in the transition metals within the
usual phonon mechanism, interest in the topic began to wane in the late
1980's. We believe that~\MgB~is a good candidate for further research on
acoustic plasmons and their contribution to superconductivity, due to its
Fermi surface characteristics. These have been first reported in
Ref.~\cite{Kortus}, and will be briefly described below:


\MgB~consists of honeycomb layers of boron atoms that are stacked vertically
with no displacement. The magnesium atoms are located at the center of the
hexagons formed by boron, but between the boron planes. The resulting
structure is called the~\chem{AlB_2}~structure. The in-plane lattice constant
is $a = \latticea$, while the interlayer spacing is $c = \latticec$. 
Hence there is a strong anisotropy in the boron-boron bond
lengths.  To a good approximation, the outer shells of the~\chem{Mg}~atoms can
be assumed as being fully ionized, so that the boron system bears a certain
resemblance with graphite, both structurally and electronically. The
differences in the electronic properties of~\MgB~and graphite are mainly due
to the attractive potential of the~\chem{Mg}~ions, which enhances the
dispersion of the $p_z$-orbitals parallel to the $c$-axis through hybridization
with the~\chem{Mg}~$s$-orbital.

\begin{figure}
  \centerline{\epsfxsize=3.0in \epsffile{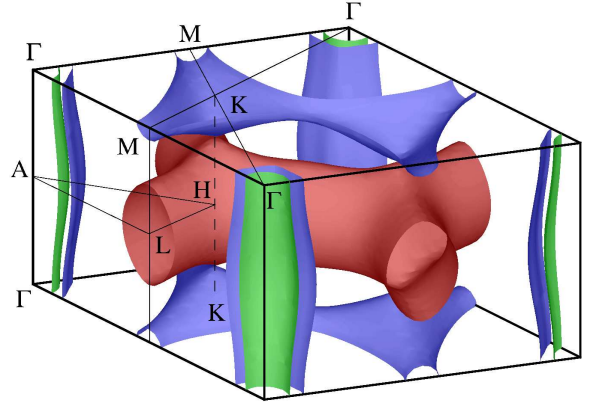}}
  \caption{Fermi surface of~\MgB. The figure is taken from
    Ref.~\protect\cite{Kortus}. Holes in the $\sigma$-band form cylinders
    around the $\Gamma A$-line. The $\pi$-band has electron and hole pockets
    located near the $H$- and $K$-points,
    respectively.\label{fig_BandStructure}}
\end{figure}

Carriers are situated in two hole bands derived from the
$\sigma$-bonding boron $p_{x,y}$-orbitals, that are essentially
two-dimensional, and in one electron and one hole band derived from
the $\pi$-bonding boron $p_z$-orbitals (see Fig.~\ref{fig_BandStructure}). The
smaller overlap of the latter in the $x,y$-directions compensates for the larger
interatomic spacing in the $z$-direction, so that the $\pi$-bands are nearly
isotropic.

An essential feature of the band structure is the large difference in the
$z$-components of the Fermi velocities between the $\sigma$ hole bands and the
$\pi$ electron and hole bands. As will be shown explicitly below, the
existence of two different velocity scales gives rise to a weakly damped
acoustic plasmon mode that corresponds to long-wavelength charge density
fluctuations between these bands. While the total charge density $\rho_\pi +
\rho_\sigma$ does not fluctuate, the charge difference $\rho_\pi -
\rho_\sigma$ will vary locally.  Since the $\sigma$-orbitals are mainly located
within the Boron planes, while the charge in the $\pi$-orbitals is 
situated between the planes, this charge transfer between the orbitals gives
rise to quadrupolar charge fluctuations.


\begin{figure}[b]
  \centerline{\epsfxsize=2.1in \epsffile{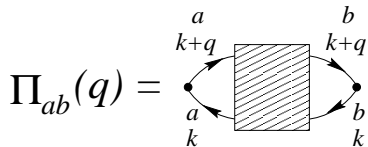}}
  \caption{Definition of the polarization insertion for different carrier
    species $a$ and $b$\label{fig_Polarization}} 
\end{figure}

We will now show the existence of a well-defined acoustic plasmon mode within
the random phase approximation (RPA). For our purpose it is convenient to
define the polarization insertion $\Pi_{ab}$ for different carrier species $a$
and $b$ as indicated in Fig.~\ref{fig_Polarization}. For a single carrier
species, $\Pi_{aa}$ then reduces to the standard definition
\cite{FetterWalecka}. The polarization insertion is related 
to the time-ordered density-density correlation function $C_{ab}$ between
different types of carriers as
\begin{equation}
  C_{ab}(\rvec,t) =  -i \left< \T \tilde n_a(\rvec,t) \tilde n_b(0,0) \right>
  = \hbar \Pi_{ab}(\rvec,t).
\end{equation}

Here $q = (\qvec,\omega)$ is a momentum four-vector, $U_0(q) = 4\pi
e^2/q^2$ is the bare Coulomb potential, and $\tilde n_a = n_a - \left< n_a
\right>$ denotes the fluctuations in the charge density of carrier species
$a$. Specializing to the case of two carrier species of opposite charge
(electrons and holes), we can express $\Pi_{ab}$ in the random phase
approximation as
\begin{equation}
\label{eq_rpa}
  \Pi_{ab}^{\rpa}(q) = \frac{ \delta_{ab} \Pi_a^0(q) - U_0(q) \Pi_1^0(q)
    \Pi_2^0(q) }{ 1 - U_0(q) \Pi_1^0(q) - U_0(q) \Pi_2^0(q) },
\end{equation}
where the lowest-order polarization insertion $\Pi_a^0$ is given in terms of
the non-interacting Green's functions $G_a^0$ as 
\begin{equation}
  \Pi_a^0(q) = \frac{2}{i\hbar} \int \frac{d^4k}{(2\pi)^4}
  \, G_a^0(k) G_a^0(k+q).
\end{equation}

We now define the total charge fluctuations $n_- = \tilde n_1 - \tilde
n_2$ and the charge transfer fluctuations $n_+ = \tilde n_1 + \tilde
n_2$. Then the corresponding correlation functions are given by
\begin{equation}
  \label{eq_cminus}
  \frac{1}{\hbar} C_-(q) = \frac{ \Pi_1^0(q) + \Pi_2^0(q) }{ 1 - 
    U_0(q)\Pi_1^0(q) - U_0(q)\Pi_2^0(q) },
\end{equation}
and
\begin{equation}
  \label{eq_cplus}
  \frac{1}{\hbar} C_+(q) = \frac{ \Pi_1^0(q) + \Pi_2^0(q) - 4U_0(q)\Pi_1^0(q)\Pi_2^0(q) }{ 1 - 
    U_0(q)\Pi_1^0(q) - U_0(q)\Pi_2^0(q) }.
\end{equation}

A collective mode corresponds to a pole in the correlation function in the
complex frequency plane, and can thus be identified by the condition
\begin{equation}
  \label{eq_plasmon_condition}
  \sum_a \Pi_a^0(q) = \frac{1}{U_0(q)},
\end{equation}
which remains valid for more than two carrier species as well.  For small wave
vectors $U_0(q)$ will be very large, so that we can set the left-hand side to
zero instead. Inserting the above condition into the
numerator of Eq.~(\ref{eq_cminus}), we see that the pole's residue is
proportional to $1/U_0(q)\sim q^2$, so that fluctuations in the total charge
density are suppressed for small wavevectors. Due to the additional term in
the numerator of Eq.~(\ref{eq_cplus}) this is not the case for the
charge transfer fluctuations described by $C_+$.


\begin{figure}[t]
  \centerline{\epsfxsize=3.4in \epsffile{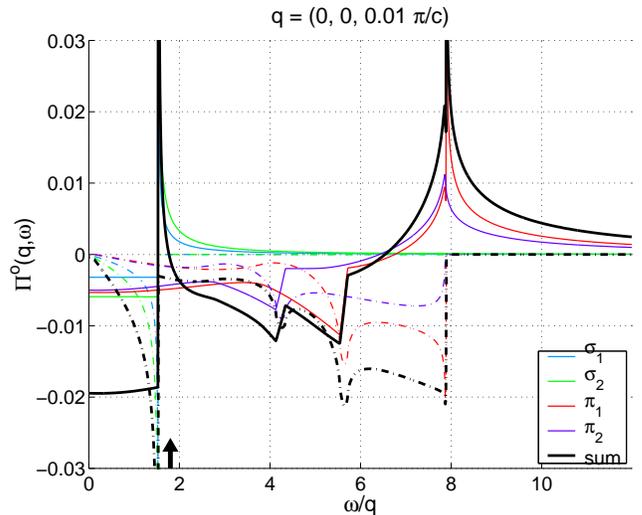}}
  \caption{Lowest-order polarization insertion $\Pi^0$ as a function of
    frequency, for a fixed wave vector in the $z$-direction. The black solid
    and dashed lines show the real and imaginary part of the total
    lowest-order polarization. The black arrow indicates the frequency of the
    acoustic plasmon mode, at which ${\mathrm{Re}}\,\Pi^0(\omega)$ vanishes.
    Also shown are the individual contributions from the two $\sigma$ hole
    bands ($\sigma_1$, blue and $\sigma_2$, green), and from the $\pi$
electron
    ($\pi_1$, red) and hole ($\Pi_2$, purple) bands. Solid lines indicate
    the real parts, and dashed lines the imaginary parts. Units for $q$ and
    $\omega$ are~$\Ang^{-1}$~and~$\eV$, respectively.
    \label{fig_Results}}
\end{figure}

We calculated the band structure with the tight-binding linear muffin-tin
orbital (TBLMTO) method \cite{LDA}, and used the following analytical fits for
our calculations of $\Pi^0$: For the $\sigma$ bands,
\begin{equation}
\label{eq_sigmabands}
  \epsilon(\kvec) = \epsilon_0^{(\sigma)} + 2 t_z^{(\sigma)} \cos(k_z c) 
  - \frac{\hbar^2}{2m_{1,2}} (k_x^2 + k_y^2),
\end{equation}
where $\epsilon_0^{(\sigma)}=0.66 \, \eV$, $t_z^{(\sigma)}=0.22 \, \eV$, $m_1
= 0.27 \, m_e$, and $m_2 = 0.50 \, m_e$, with $m_e$ being the bare electron
mass. For the $\pi$ bands we used
\begin{eqnarray}
  && \epsilon(\kvec) = \epsilon_0^{(\pi)} - 2 t_z^{(\pi)} \cos(k_z c) \\
  && \pm 2 t_x \sqrt{\frac{3}{4} + \frac{1}{2} \cos(k_x a) 
    + \cos(\frac{1}{2}k_x a) \cos(\frac{\sqrt{3}}{2}k_y a)}, \nonumber
\end{eqnarray}
where $\epsilon_0^{(\pi)}=0.16 \, \eV$, $t_z^{(\pi)}=1.12 \, \eV$, $t_x=1.73
\, \eV$, and the positive or negative sign corresponds to the electron and
hole band, respectively.

Within the approximate dispersion given by Eq.~(\ref{eq_sigmabands}), the
lowest-order polarization insertion of the $\sigma$ bands can be calculated
exactly for the special case where $\qvec$ is parallel to the $z$-axis:
\begin{equation}
  \label{eq_analytic}
  \Pi^0_{\sigma1,2}(q \hat z,\omega) = 
  \frac{m_{1,2}}{\pi \hbar^2 c} 
  \left\{ \frac{ \theta(1-Y^2) - i \, \theta(Y^2-1) }
    { \sqrt{ \left| 1 - Y^2 \right| } } - 1 \right\},
\end{equation}
where $Y=4t_z^{(\sigma)}\sin(qc/2)/\hbar\omega$, and $\theta(x)$ is the
Heaviside step function.  For $\qvec$ off the z-axis and for the $\pi$-bands
we resort to numerical calculations. Results are shown in
Fig.~\ref{fig_Results}. From the condition (\ref{eq_plasmon_condition}) we can
identify an acoustic plasmon mode with dispersion $\omega_a =\va q$, where the
plasmon velocity is $\va \simeq 1.78 \, \mathrm{\AA \, eV} = 2.70 \times 10^7
\, \mathrm{{cm}/{s}}$.  The acoustic plasmon mode is weakly damped by decay
into the $\pi$ electron-hole continuum.  The remaining two zeroes of
${\mathrm{Re}} \, \Pi^0$ lie deep within the electron-hole continuum and
correspond to overdamped modes, as can be seen from the large imaginary part
of $\Pi^0$.  To estimate the damping coefficient $\lambda$ of the weakly
damped mode we assume that the contribution from the $\pi$ carriers to the
real and imaginary parts of $\Pi^0$ is constant in the vicinity of $\omega_a$.
Then the imaginary part of the pole's location in the complex plane can easily
be calculated from the analytic expression~(\ref{eq_analytic}), and yields a
Q-factor $Q=\omega_a/\lambda\simeq 15$. The acoustic plasmon mode is hence
well-defined.


The above analysis leads us to propose that a well defined acoustic plasmon
mode exists for small values of the transverse component $q_\perp$. Physically
this mode corresponds to quadrupolar charge fluctuations involving a local
charge transfer between $\sigma$- and $\pi$-orbitals of the boron atoms. This
collective mode in turn leads to an attractive retarded contribution to the
Coulomb interaction in a window in $(\qvec,\omega)$-space, which will modify
the repulsive Coulomb pseudopotential. Further analysis is required to
determine if the modification is strong enough to give the proposed reduction
to a value of $\mu^*=0.02$.


\begin{references}
\bibitem{Akimitsu} J.~Nagamatsu, N.~Nakagawa, T.~Muranaka, and J.~Akimitsu, 
  Nature \textbf{410}, 63 (2001).
\bibitem{Andersen} Y.~Kong, O.~V.~Dolgov, O.~Jepsen, and O.~K.~Andersen,
  preprint \textit{cond-mat/0102499} (2001).
\bibitem{Pines} D.~Pines, Can.~J.~Phys.~\textbf{34}, 1379 (1956).
\bibitem{Froehlich} H.~Fr\"ohlich, J.~Phys.~C \textbf{1}, 545 (1986).
\bibitem{Kortus} J.~Kortus, I.~I.~Mazin, L.~D.~Belashchenko, V.~P.~Antropov,
  and L.~L.~Boyer, preprint \textit{cond-mat/0101446} (2001).
\bibitem{Budko} S.~L.~Bud'ko, G.~Lapertot, C.~Petrovic, C.~E.~Cunningham,
  N.~Anderson, and P.~C.~Canfield, Phys.~Rev.~Lett.~\textbf{86}, 1877 (2001). 
\bibitem{FetterWalecka} A.~L.~Fetter and J.~D.~Walecka, \textit{Quantum Theory 
    of Many-Particle Systems}, McGraw-Hill, 1971.
\bibitem{LDA} O.~K.~Andersen, Phys.~Rev.~B \textbf{12}, 3060 (1975);
  O.~K.~Andersen, Z.~Pawlowska, and O.~Jepsen, Phys.~Rev.~B \textbf{34}, 5253 
  (1984).

\end{references}
\end{document}